\documentstyle[psfig,bm]{nature}

\newcommand{\bsf}{\sffamily\bfseries}

\newcommand{\msun}{\mbox{M$_{\odot}$}}

\newcommand{\psr}{\mbox{PSR J0437$-$4715}}

\newcommand\wvect[1]{{\bsf #1}}
\newcommand\bvect[1]{{\bsf #1}$_{\bm 0}$}

\newcommand\fs{\mbox{$.\!\!^{\mathrm s}$}}%
\newcommand\arcmin{\mbox{$^\prime$}}%
\newcommand\farcs{\mbox{$.\!\!^{\prime\prime}$}}%
\newcommand\fdegr{\mbox{$.\!\!^{\circ}$}}%
\newcommand\etal{et al.}

\def\degr{\ifmmode^{\circ}\else$^{\circ}$\fi}
\def\agep{\ifmmode\tau_{\rm c}\else$\tau_{\rm c}$\fi}
\def\agec{\ifmmode\tau_{\rm w}\else$\tau_{\rm w}$\fi}

\title{A test of general relativity from \\
	the three-dimensional orbital geometry of a binary pulsar }

\newcommand{\sut}{Centre for Astrophysics and Supercomputing, Swinburne
University of Technology, P.O.~Box 218 Hawthorn, VIC 3122, Australia}

\newcommand{\cit}{Division of Physics, Mathematics, and Astronomy,
California Institute of Technology, Mail Code 220-47,
Pasadena, California 91125 }

\newcommand{\atnf}{Australia Telescope National Facility -- CSIRO,
P.O. Box 76, Epping NSW 1710, Australia}

\author{ W. van Straten\affiliation{\sut}, M. Bailes$^*$,
	M.C. Britton$^*$, \\
	S.R. Kulkarni\affiliation{\cit}, S.B. Anderson$^\dagger$,
	R.N. Manchester\affiliation{\atnf} \& J. Sarkissian$^\ddag$}
 
\mainauthor{W. van Straten \etal}
\headertitle { Three-Dimensional Orbital Geometry and Shapiro Delay of \psr }

\summary{
Binary pulsars provide an excellent system for testing general
relativity because of their intrinsic rotational stability and the
precision with which radio observations can be used to determine their
orbital dynamics.  Measurements of the rate of orbital decay of two
pulsars have been shown to be consistent with the emission of
gravitational waves as predicted by general
relativity\cite{tw82,sac+98}, providing the most convincing evidence
for the self-consistency of the theory to date.  However, independent
verification of the orbital geometry in these systems was not
possible.  Such verification may be obtained by determining the
orientation of a binary pulsar system using only classical geometric
constraints, permitting an independent prediction of general
relativistic effects.  Here we report high-precision timing of the
nearby binary millisecond pulsar \psr, which establish the
three-dimensional structure of its orbit.  We see the expected
retardation of the pulse signal arising from the curvature of
space-time in the vicinity of the companion object (the `Shapiro
delay'), and we determine the mass of the pulsar and its white dwarf
companion.  Such mass determinations contribute to our understanding
of the origin and evolution of neutron stars\cite{tc99}.  }

\begin{document}
\maketitle

Discovered in the Parkes \mbox{70-cm survey\cite{jlh+93}}, \psr\ remains the
closest and brightest millisecond pulsar known.  It is bound to a
low-mass helium white dwarf companion\cite{bbb93,dbd93} in a nearly
circular orbit.  Owing to its proximity, relative motion between the
binary system and the Earth significantly alters the line-of-sight
direction to the pulsar and, consequently, the orientation of the
basis vectors used in the timing model (see \mbox{Fig. 1}).  Although
the physical orientation of the orbit in space remains constant, its
parameters are measured with respect to this time-dependent basis and
therefore also vary with time.  Variations of the inclination angle,
$i$, change the projection of the semi-major axis along the
line-of-sight, $x\equiv a_{\rm p}\sin i/c$, where $a_{\rm p}$ is the
semi-major axis of the pulsar orbit.

The heliocentric motion of the Earth induces a periodic variation
of $x$ known as the annual-orbital parallax\cite{kop95},
\begin{equation}
x^{\rm obs}(t)=x^{\rm int}[1+{\cot i\over d}\ 
	{\bf r_\oplus}(t)\cdot\bm{\Omega^\prime}].
\label{eqn:aopx}
\end{equation}
The superscripts `obs' and `int' refer to the observed and intrinsic
values, respectively, ${\bf r_\oplus}(t)$ is the position vector of
the Earth with respect to the barycentre of the Solar System as a
function of time, $d$ is the distance to the pulsar, and
$\bm{\Omega^\prime}=\sin\Omega{\bf I_0} - \cos\Omega{\bf J_0}$ (see
\mbox{Fig. 1}).  Similarly, the proper motion of the
binary system induces secular evolution of the projected semi-major
axis\cite{kop96,sbm+97}, such that:
\begin{equation}
\dot x^{\rm obs}=\dot x^{\rm int}-x\cot i\ \bm{\mu}\cdot\bm{\Omega^\prime};
\label{eqn:xdot}
\end{equation}
where $\bm{\mu}=\mu_\alpha\bf{I_0} + \mu_\delta\bf{J_0}$ is the proper
motion vector with components in right ascension, $\mu_\alpha$, and
declination, $\mu_\delta$.  An apparent transverse quadratic Doppler
effect (known as the Shklovskii effect) also arises from the system's
proper motion and contributes to the observed orbital period
derivative\cite{dt91}:
\begin{equation}
\dot P_{\rm b}^{\rm obs}=\dot P_{\rm b}^{\rm int} + \beta P_{\rm b};
\label{eqn:pbdot}
\end{equation}
where $\beta=\mu^2d/c$, and $\mu=|\bm{\mu}|$.

Observations of \psr\ were conducted from 11 July 1997 to 13 December
2000, using the Parkes 64\,m radio telescope.  Over 50 terabytes of
baseband data have been recorded with the S2 Recorder\cite{cbf+97} and
the Caltech Parkes Swinburne Recorder (CPSR)\cite{vbb00}, followed by
offline reduction at Swinburne's supercomputing facilities.  Average
pulse profiles from hour-long integrations were fitted to a high
signal-to-noise template\cite{tay92}, producing a total of 617 pulse
arrival time measurements with estimated errors on the order of 100
ns.

Previously considered negligible, the annual-orbital parallax has been
largely ignored in experimental time-of-arrival analyses to date.
However, our initial estimates of its peak-to-peak amplitude for \psr\
\mbox{($\sim$ 400 ns)} demonstrated that it would be clearly
detectable above the timing noise.  As can be seen in \mbox{equation
\ref{eqn:aopx}}, $x^{\rm obs}$ varies with a period of one year and
phase determined by $\bm{\Omega^\prime}$.  Its inclusion in our timing
model therefore provides a geometric constraint on $\Omega$.  We also
note that the value of $\dot x^{\rm obs}=(7.88\pm0.01)\times10^{-14}$
observed in our preliminary studies is many orders of magnitude larger
than the intrinsic $\dot x$ expected as a result of the emission of
gravitational waves, $\dot x^{\rm GR}=-1.6\times10^{-21}$.  Neglecting
$\dot x^{\rm int}$, the relationship between $i$ and $\Omega$ defined
by \mbox{equation \ref{eqn:xdot}} is parameterized by the well
determined physical parameters $x$, $\dot x$ and $\bm{\mu}$.  Also,
because $\bm{\mu}$ is fortuitously nearly anti-parallel to
$\bm{\Omega^\prime}$, $\delta i$/$\delta \Omega$ is close to zero, and
incorporation of \mbox{equation \ref{eqn:xdot}} in our timing model
provides a highly significant constraint on the inclination angle.

\begin{figure}[h]
\centerline{\psfig{figure=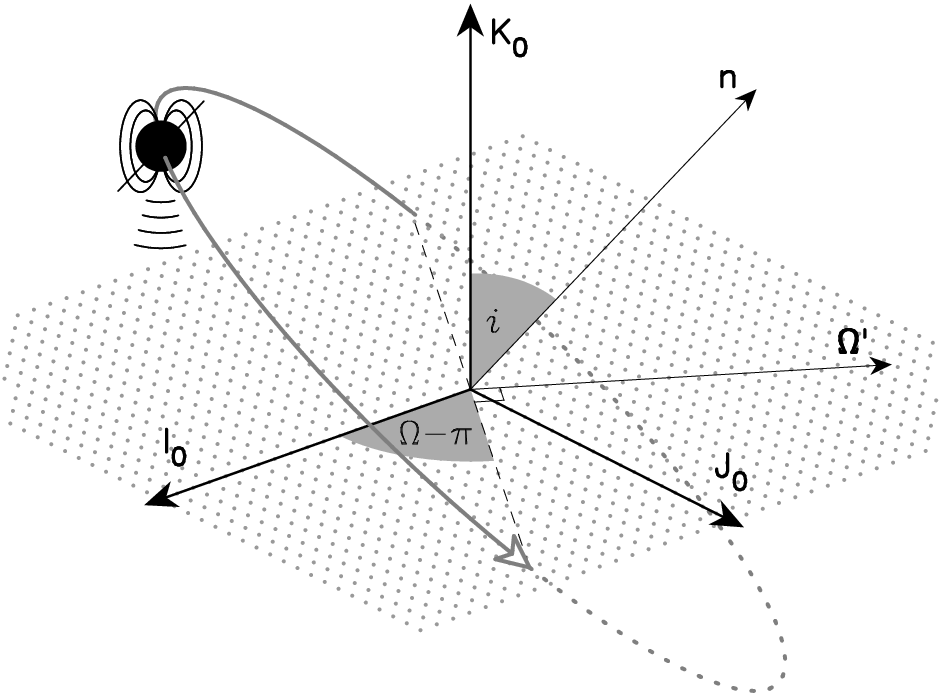,width=80mm,angle=-90}}
\label{fig:geometry}
{\sffamily {\bfseries Figure 1} The three-dimensional orientation of
the pulsar orbit is determined using a classical geometric model.
With the centre of mass of the binary system at the origin, the basis
vectors, \bvect{I}, \bvect{J} and \bvect{K}, define east, north, and
the line-of-sight from Earth, respectively.  The orientation of the
normal vector, \wvect{n}, is defined with respect to this basis by the
longitude of the ascending node, $\Omega$, and the inclination angle,
$i$.  The plane of the sky, or \bvect{I}--\bvect{J} plane, (shown
stippled) intersects the orbital plane at the ``line of nodes''
(dashed line).  Below the \bvect{I}--\bvect{J} plane, the orbital path
has been drawn with a dotted line.  The unit vector,
$\bm{\Omega^\prime}$, lies in the \bvect{I}--\bvect{J} plane and is
perpendicular to the line of nodes.  The pulsar is shown at superior
conjunction, where radio pulses emitted toward Earth experience the
greatest time delay due to the gravitating mass of the companion on
the opposite side of the centre of mass. }
\end{figure}

The orbital inclination parameterizes the shape of the Shapiro delay,
that is, the delay due to the curvature of space-time about the
companion.  In highly inclined orbits, seen more edge-on from Earth,
the companion passes closer to the line-of-sight between the pulsar
and the observatory, and the effect is intensified.  As the relative
positions of the pulsar and companion change with binary phase, the
Shapiro delay also varies and, in systems with small orbital
eccentricity, is given by:
\begin{equation}
\Delta_{\rm S}=-2r\ln[1-s\cos(\phi-\phi_0)].
\label{eqn:shapiro}
\end{equation}
Here, $s\equiv\sin i$ and $r\equiv Gm_2/c^3$ are the shape and range,
respectively, $\phi$ is the orbital phase in radians, and $\phi_0$ is
the phase of superior conjunction, where the pulsar is on the opposite
side of the companion from Earth (as shown in \mbox{Fig. 1}).  For
small inclinations, the orbit is seen more face-on from Earth, and
$\Delta_{\rm S}$ becomes nearly sinusoidal in form.

In the \psr\ system, the Shapiro effect is six orders of magnitude
smaller than the classical Roemer delay, the time required for light
to travel across the pulsar orbit.  In nearly circular orbits, the
Roemer delay also varies sinusoidally with binary phase.
Consequently, when modeling less inclined binary systems with small
eccentricity, the Shapiro delay can be readily absorbed in the Roemer
delay by variation of the classical orbital parameters, such as $x$.
For this reason, a previous attempt at measuring the Shapiro effect in
the \mbox{PSR J1713+0747} system\cite{cfw94} yielded only weak,
one-sided limits on its shape and range.

In contrast, we have significantly constrained the shape independently
of general relativity, enabling calculation of the component of
$\Delta_{\rm S}$ that remains un-absorbed by the Roemer delay.  The
theoretical signature is plotted in \mbox{Fig. 2} against post-fit
residuals obtained after fitting the arrival time data to a model that
omits the Shapiro effect.  To our knowledge, this verification of the
predicted space-time distortion near the companion is the first such
confirmation (outside our Solar System) in which the orbital
inclination was determined independently of general relativity.

\begin{figure}[h]
\centerline{\psfig{figure=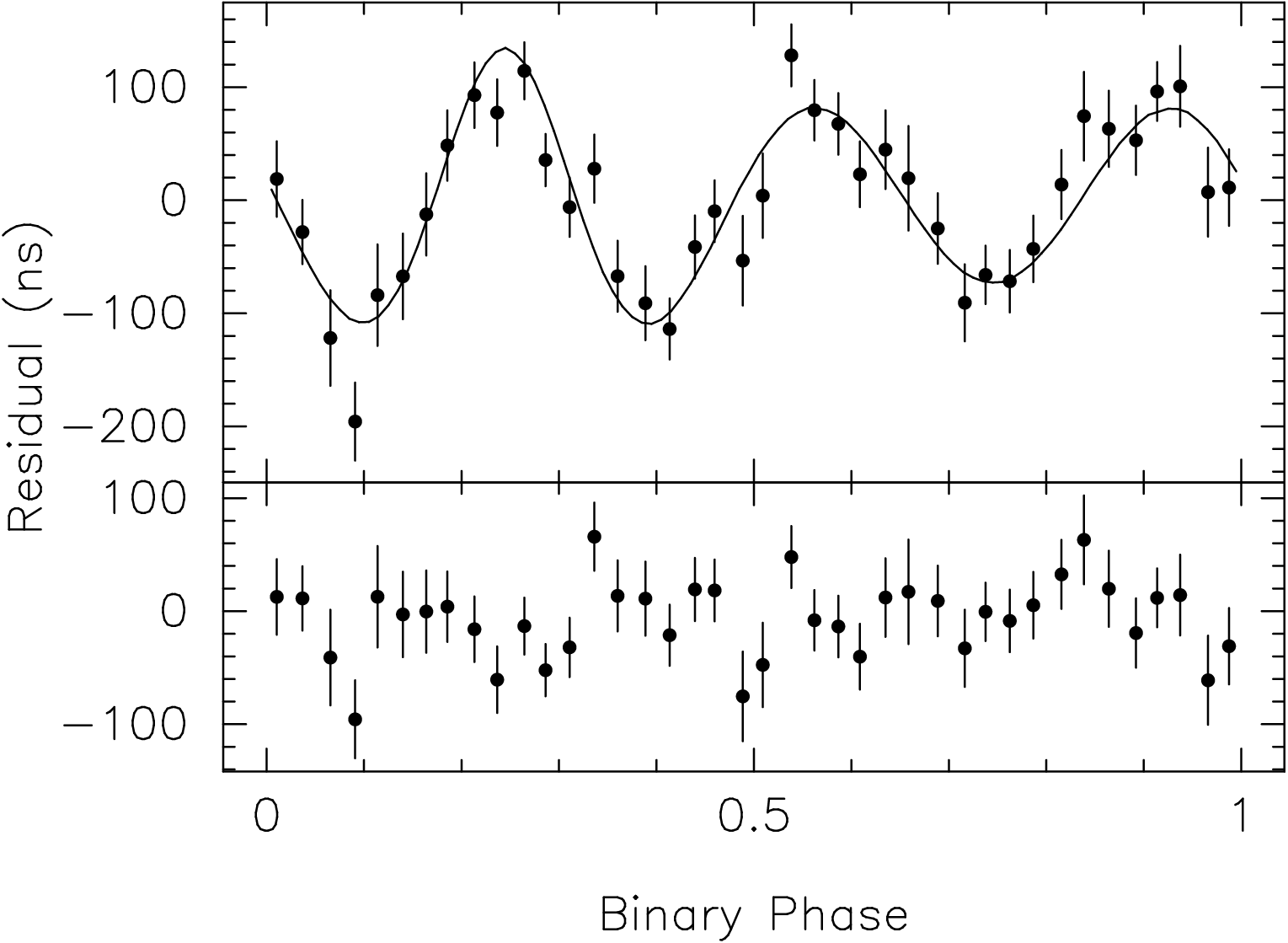,width=80mm,angle=-90}}
\label{fig:shapiro}
{ \sffamily {\bfseries Figure 2} Arrival time residuals confirm
the predicted space-time distortion induced by the pulsar
companion.  The unabsorbed remnant of Shapiro Delay is much smaller
than the theoretical total delay, which for \psr\ has a peak-to-peak
amplitude of about \mbox{3.8\,$\mu$s}.  In the top panel, the solid line
models the expected delay resulting from a companion with a mass of
\mbox{0.236 \msun}, at the geometrically-determined orbital
inclination.  Measured arrival time residuals, averaged in 40 binary
phase bins and plotted with their 1$\sigma$ errors, clearly exhibit
the predicted signature.  In the bottom panel, the same residuals with
the model removed have an r.m.s. residual of only 35 ns and
a reduced $\chi^2$ of 1.13. }
\end{figure}

\noindent
The range of the Shapiro delay provides an estimate of the companion
mass, $m_2=0.236\pm0.017$~\msun, where \msun is the mass of the
sun.  Through the mass function\cite{tc99}, $f(M)$, we then obtain a
measurement of the pulsar mass \mbox{$m_p=1.58\pm0.18$ \msun}.
Slightly heavier than the proposed average neutron star
mass\cite{tc99}, \mbox{$m_{\rm p}^{\rm avg}=1.35\pm0.04$ \msun}, this
value of $m_p$ suggests an evolutionary scenario that includes an
extended period of mass and angular momentum transfer.  Such accretion
is believed to be necessary for a neutron star to attain a spin period
of the order of a millisecond\cite{tv86}.  It is also expected that,
during accretion, the pulsar spin and orbital angular momentum vectors
are aligned.  Under this assumption, the measured inclination angle of
$i=42\fdegr75\pm0\fdegr09$ does not support the conjecture that pulsar
radiation may be preferentially beamed in the equatorial
plane\cite{bac98}.

The total system mass, $M$, can also be calculated from the observed
$\dot\omega$, using the general relativistic prediction of the rate of
orbital precession.  Using $M$, $f(M)$, and $i$, we obtain a second
consistent estimate of the companion mass, $m_2^\prime=0.23\pm0.14$
\msun, the precision of which is expected to increase with time as
$t^{3/2}$, surpassing that of the $r$-derived value in approximately
30 years.

The complete list of physical parameters modelled in our analysis is
included in \mbox{Table 1}.  Most notably, the pulsar position,
parallax distance, $d_\pi=139\pm3$ pc, and proper motion,
\mbox{$\mu=140.892\pm0.006$ mas yr$^{-1}$}, are known to accuracies
unsurpassed in astrometry.  Although closer, $d_\pi$ lies within the
1.5 $\sigma$ error of an earlier measurement by \scite{sbm+97},
\mbox{$178\pm26$ pc}.  The $d_\pi$ and $\mu$ estimates can be used to
calculate $\beta$ and the intrinsic spin period derivative, $\dot
P^{\rm int}=\dot P^{\rm obs}-\beta P=(1.86\pm0.08)\times10^{-20}$,
providing an improved characteristic age of the pulsar,
\mbox{\agep$=P/(2\dot P^{\rm int})=4.9$ Gyr}.
Another distance estimate may be calculated using the
observed $\mu$ and $\dot P_{\rm b}$ by solving Equation
\ref{eqn:pbdot} for $d$, after noting the relative negligibility of
any intrinsic contribution\cite{bb96}. The precision of the derived
value, \mbox{$d_{\rm B}=150\pm9$ pc}, is anticipated to improve as
$t^{5/2}$, providing an independent distance estimate with relative
error of about 1\% within the next three to four years.


With a post-fit root mean square (r.m.s.) residual of merely 130 ns
over 40 months, the accuracy of our analysis has enabled the detection
of annual-orbital parallax.  This has yielded a three-dimensional
description of a pulsar binary system and a new geometric verification
of the general relativistic Shapiro delay.  Only the Space
Interferometry Mission (SIM) is expected to localize celestial objects
with precision similar to that obtained for \psr\ (including
parallax). By the time SIM is launched in 2010, the precision of this
pulsar's astrometric and orbital parameters will be vastly improved.
Observations of the companion of \psr\ using SIM will provide an
independent validation and a tie between the SIM frame and the
solar-system dynamic reference frame.

We also expect that continued observation and study of this pulsar
will ultimately have an important impact in cosmology.  Various
statistical procedures have been applied to the unmodelled residuals of
\mbox{PSR B1855+09} (see ref. 19 and references therein) in an 
effort to place a rigorous upper limit on $\Omega_g$, the fractional
energy density per logarithmic frequency interval of the primordial
gravitational wave background.  As the timing baseline for \psr\
increases, our experiment will probe more deeply into the low
frequencies of the cosmic gravitational wave spectrum, where, owing to
its steep power-law dependence\cite{mzvl96}, the most stringent
restriction on $\Omega_g$ can be made.

\bibliographystyle{nature}
\bibliography{modrefs,psrrefs,journals,../local}

\vspace{5mm}
\noindent
\bsf{Acknowledgements} \\

\noindent
{\small The Parkes Observatory is part of the Australia Telescope
which is funded by the Commonwealth of Australia for operation as a
National Facility managed by CSIRO.  We thank the staff at Parkes
Observatory for technical assistance and performance of regular
observations.  S.R.K. and S.B.A. thank NSF and NASA for supporting
their work at Parkes.  We also thank R. Edwards and M. Toscano for
comments on the text.  We received support from Compaq and the Space
Geodynamics Laboratory of the Centre for Research in Earth and Space
Technology.  M.~Bailes is an ARC Senior Research Fellow. \\
}

\begin{table}
\sffamily
\begin{center}
\begin{tabular}{lr}
\hline
\multicolumn{2}{c}{\bf\textsf{Table 1 PSR J0437--4715 physical parameters}} \\
\hline \\
Right ascension, $\alpha$ (J2000)       \dotfill &
			04$^{\mathrm h}$37$^{\mathrm m}$15\fs7865145(7) \\
Declination, $\delta$ (J2000)           \dotfill & 
			-47\degr15\arcmin08\farcs461584(8) \\
$\mu_\alpha$ (mas yr$^{-1}$) \dotfill & 121.438(6) \\
$\mu_\delta$ (mas yr$^{-1}$) \dotfill & -71.438(7) \\
Annual parallax, $\pi$ (mas)              \dotfill & 7.19(14) \\
Pulse period, $P$ (ms)                    \dotfill & 5.757451831072007(8) \\
Reference epoch (MJD)                     \dotfill & 51194.0  \\
Period derivative, $\dot{P}$ (10$^{-20}$) \dotfill & 5.72906(5) \\
Orbital period, $P_{\rm b}$ (days)        \dotfill & 5.741046(3) \\
$x$ (s)        \dotfill & 3.36669157(14) \\
Orbital eccentricity, $e$        	  \dotfill & 0.000019186(5) \\
Epoch of periastron, $T_{0}$ (MJD)        \dotfill & 51194.6239(8) \\
Longitude of periastron, $\omega$ (\degr) \dotfill & 1.20(5) \\
Longitude of ascension, $\Omega$ (\degr)  \dotfill & 238(4) \\
Orbital inclination, $i$ (\degr)          \dotfill & 42.75(9) \\
Companion mass, $m_2$ (M$_{\odot})$       \dotfill & 0.236(17) \\
$\dot P_{\rm b} (10^{-12})$      	  \dotfill & 3.64(20) \\
$\dot \omega$ (\degr yr$^{-1}$)           \dotfill & 0.016(10) \\
\end{tabular} \\

\vspace{5mm}
\parbox{88mm}{\sffamily Best-fit physical parameters and their formal 1$\sigma$
errors were derived from arrival time data by minimizing an objective
function, $\chi^2$, as implemented in TEMPO
(http://pulsar.princeton.edu/tempo).  Our timing model is based on the
relativistic binary model\cite{dd86} and incorporates additional
geometric constraints derived by Kopeikin\cite{kop95,kop96}.
Indicative of the solution's validity, $\chi^2$ was reduced by 30\%
with the addition of only one new parameter, $\Omega$.  To determine
the 1$\sigma$ confidence intervals of $\Omega$ and $i$, we mapped
projections of the $\Delta\chi^2\equiv\chi^2(\Omega,i)-\chi^2_{\rm
min}=1$ contour, where $\chi^2(\Omega,i)$ is the value of $\chi^2$
minimized by variation of the remaining model parameters, given
constant \mbox{$\Omega$ and $i$}. Parenthesized numbers represent
uncertainty in the last digits quoted, and epochs are specified using
the Modified Julian Day (MJD).}
\end{center}
\end{table}

\end{document}